\documentclass[superscriptaddress, reprint]{revtex4-2}
\usepackage{multirow}
\usepackage{hyperref}
\usepackage[version=4]{mhchem} 
\usepackage{graphicx}
\usepackage{chemmacros}
\usepackage{gensymb}
\usepackage{multirow}
\usepackage{dcolumn}
\usepackage{amssymb}
\usepackage{xspace}

\newcommand{\koneqq} {\ensuremath{\textbf{k}_{\mathrm{qq}} }}
\newcommand{\koneqmq}{\ensuremath{\mathbf{k}_{\mathrm{q\bar{q}}} }}
\newcommand{\ktwozq} {\ensuremath{\mathbf{k}^2_{\mathrm{q0}} }}
\newcommand{\ktwoqz}{\ensuremath{\mathbf{k}^2_{\mathrm{0q}} }}
\newcommand{\kthreezq}{\ensuremath{\mathbf{k}^3_{\mathrm{0q}} }}
\newcommand{\kfourzq}{\ensuremath{\mathbf{k}^4_{\mathrm{0q}} }}

\begin{document}

\title{Spontaneous spin chirality reversal and competing phases in the topological magnet \ce{EuAl4}}

\author{A. M. Vibhakar}
\email{anuradha.vibhakar@diamond.ac.uk}
\affiliation{Diamond Light Source Ltd,  Harwell Science and Innovation Campus, Didcot, Oxfordshire, OX11 0DE, United Kingdom}
\author{D. D. Khalyavin}
\affiliation{ISIS facility, Rutherford Appleton Laboratory-STFC, Chilton, Didcot, OX11 0QX, United Kingdom}
\author{F. Orlandi}
\affiliation{ISIS facility, Rutherford Appleton Laboratory-STFC, Chilton, Didcot, OX11 0QX, United Kingdom}
\author{J. M. Moya}
\affiliation{Applied Physics Graduate Program, Smalley-Curl Institute, Rice University, Houston, Texas 77005, USA}
\affiliation{Department of Physics and Astronomy, Rice University, Houston, Texas, 77005 USA}
\altaffiliation{Current address: Department of Chemistry, Princeton University, Princeton, New Jersey 08544, USA}
\author{S. Lei}
\affiliation{Department of Physics and Astronomy, Rice University, Houston, Texas, 77005 USA}
\affiliation{Hong Kong University of Science and Technology, Clear Water Bay, Hong Kong}
\author{E. Morosan}
\affiliation{Department of Physics and Astronomy, Rice University, Houston, Texas, 77005 USA}
\author{A. Bombardi}
\email{alessandro.bombardi@diamond.ac.uk}
\affiliation{Diamond Light Source Ltd,  Harwell Science and Innovation Campus, Didcot, Oxfordshire, OX11 0DE, United Kingdom}

\date{\today}

\begin{abstract}
We demonstrate the spontaneous reversal of spin chirality in a single crystal sample of the intermetallic magnet \ce{EuAl4}. We solve the nanoscopic nature of each of the four magnetically phases of \ce{EuAl4} using resonant magnetic x-ray scattering, and demonstrate all four phases order with single-k incommensurate magnetic modulation vectors. Below 15.4 K the system forms a spin density modulated spin structure where the spins are orientated in the $ab$ plane perpendicular to the orientation of the magnetic propagation vector. Below 13.2 K a second spin density wave orders with moments aligned parallel to the $c$-axis, such that the two spin density wave orders coexist. Below 12.2 K a magnetic helix of a single chirality is stabilised across the entire sample. Below 10 K the chirality of the magnetic helix reverses, and the sample remains a single chiral domain. Concomitant with the establishment of the helical magnetic ordering is the lowering of the crystal symmetry to monoclinic, as evidenced the formation of uniaxial charge and spin strip domains. A group theoretical analysis demonstrates that below 12.2 K the symmetry lowers to polar monoclinic, which is necessary to explain the observed asymmetry in the chiral states of the magnetic helix and the spin chiral reversal. We find that in every magnetically ordered phase of \ce{EuAl4} the in-plane moment is perpendicular to the orientation of the magnetic propagation vector, which we demonstrate is favoured by magnetic dipolar interactions.

\end{abstract}

\maketitle

\section{Introduction}

The bivalent nature of chirality, i.e a chiral system is either left or right handed, lends itself well to the encoding of binary information for classical computation. This is reflected in a surge in interest to control the chiral magnetic states of a material through electronic and spin degrees of freedom, for instance via non-collinear or polarised spin currents \cite{2021Yang, 2022Go, 2023Kondou, 2024Lee}. Chiral magnetism can occur in inversion symmetric crystals, for which competing exchange interactions are suggested to play a role in stabilising the chiral order \cite{2015Leonov, 2021Hayami, 2022Takagi}, or in non-centrosymmetric crystals that allow for Dzylaoshinskii-Moriya exchange interaction, which gives rise to the spin chirality \cite{2008Marty, 2010Janoschek, 2022Cheong}. In the latter case it is often the coupling between the macroscopic order parameter that breaks the inversion symmetry, and spin that allows for the field controlled reversal of the chirality \cite{2022Behera}. Reversing the spin chirality in the absence of an applied field could have far reaching consequences for the design and development of future spintronic devices, just as materials that exhibit a spontaneous rotation of the magnetisation have had \cite{2014Locatelli, 2018Fita, 2020Vibhakar}. 

Besides the complex and intriguing zero field magnetism, demonstrated by a cascade of four different phase transitions in a temperature interval of only 5 K, \ce{EuAl4} is also shown to stabilise seven different magnetic phases and four different skyrmion lattices under an applied magnetic field \cite{2015Nakamura, 2022Meier, 2022Takagi}. \ce{EuAl4} is expected to be centrosymmetric, and hence the existence of chiral topological objects such as skymrions, was explained not by the Dzyaloshinskii Moriya interaction, but Ruderman-Kittel-Kasuya-Yosida (RKKY) and frustrated itinerant interactions \cite{2022Hayami, 2022Takagi}. 

\ce{EuAl4} crystallises in the tetragonal space group $I4/mmm$ at room temperature. The Eu$^{2+}$ ions, Wyckoff position $2a$, form two-dimensional square layers in the $ab$ plane. Neighbouring Eu layers, which are separated along $c$, are related by the I-centering translation \cite{2018Stavinoha}. We label the Eu ion at the vertex Eu11, and the Eu ion at the body centre as Eu12. Below T$_{\mathrm{CDW}}$ = 145 K the onset of charge density wave order occurs  with propagation vector (0,0,$\tau$) $\tau$ = 0.1781(3) \cite{2014Araki, 2019Shimomura, 2022Ramakrishan}. Single crystal neutron diffraction studies demonstrated that the four magnetically ordered phases of \ce{EuAl4}, which we label as AFM1, AFM2, AFM3 and AFM4 with onset temperatures of T$_1$ = 15.4 K, T$_2$ = 13.2 K, T$_3$ = 12.2 K and T$_4$ $\sim$ 10 K respectively, ordered with multiple magnetic propagation vectors \cite{2021Kaneko}. However the nature of the magnetic phases remained an open question, as did the physical mechanisms driving the changing modulations. 

In this manuscript we demonstrate the spontaneous reversal of spin chirality. We solve the nanoscopic nature of each of the four magnetically phases of \ce{EuAl4} using resonant magnetic x-ray scattering. We show that all four phases are single-k, where the AFM1 and AFM2 phases are characterised by spin density wave order, and the AFM3 and AFM4 phases by magnetic helices of a single chiral domain. In the AFM4 phase, the nature of the magnetic ordering does not change, a magnetic helix of a single chirality is still present, but its chirality reverses. We find that in all four magnetically ordered phases the in-plane component of the spins are orientated perpendicular to the direction of the propagation vector, which we demonstrate is favoured by magnetic dipolar interactions. Using a group theoretical approach to symmetry analysis we demonstrate that at the onset of the helical magnetic ordering below T$_3$, the symmetry lowers to polar monoclinic, which explains the observed asymmetry in the chiral states of the magnetic helix.

\begin{figure}[ht]
\centering
\includegraphics[width =\linewidth]{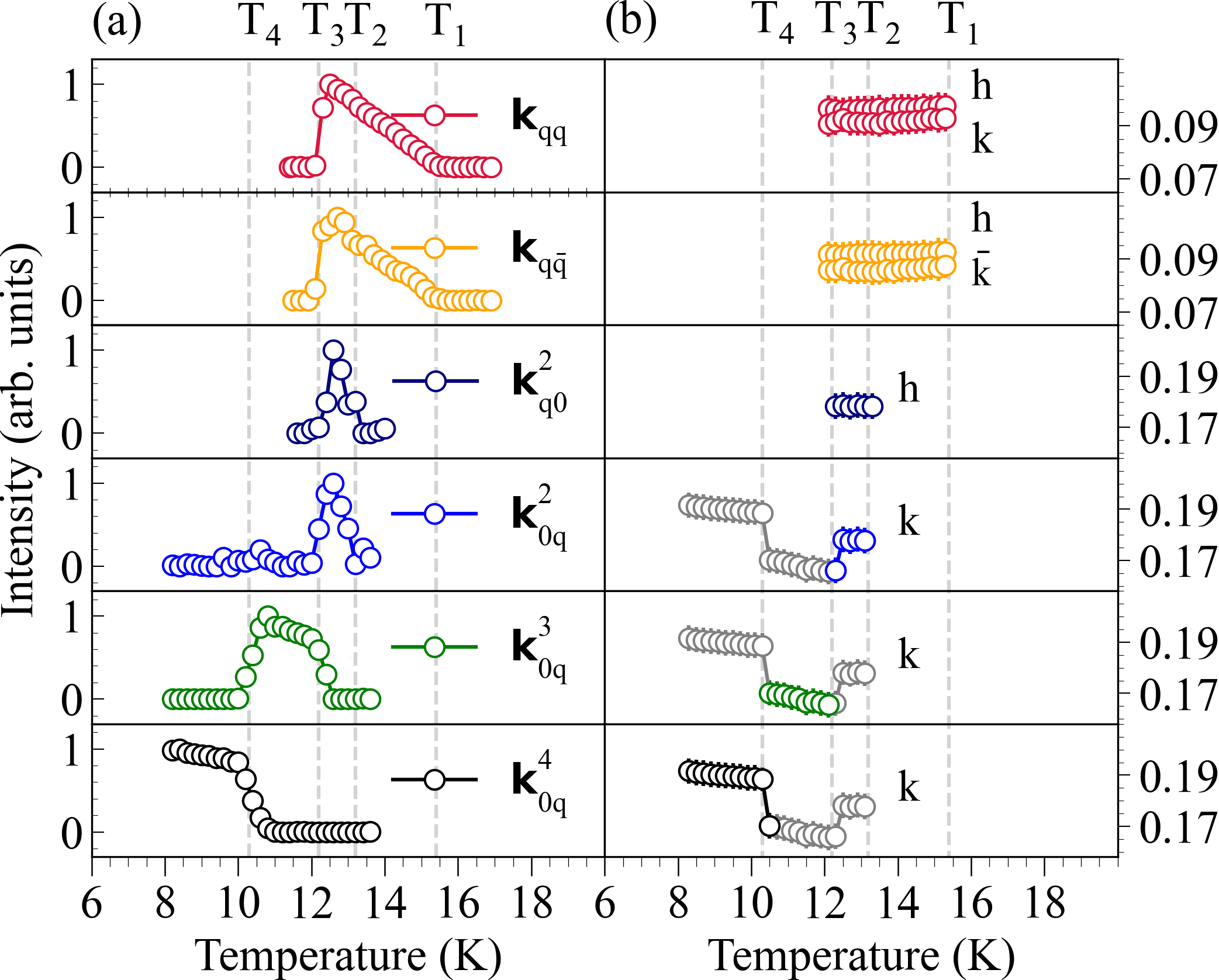}
\caption{\label{FIG::tempdep} Temperature dependence of (a) the integrated intensity and (b) the magnitude of the propagation vector along the h and k directions for the \koneqq , \koneqmq , \ktwozq , \ktwoqz , \kthreezq, \kfourzq magnetic satellites of the (0,0,8) reflection.}
\end{figure}

\begin{figure*}[ht]
\centering
\includegraphics[width =\linewidth]{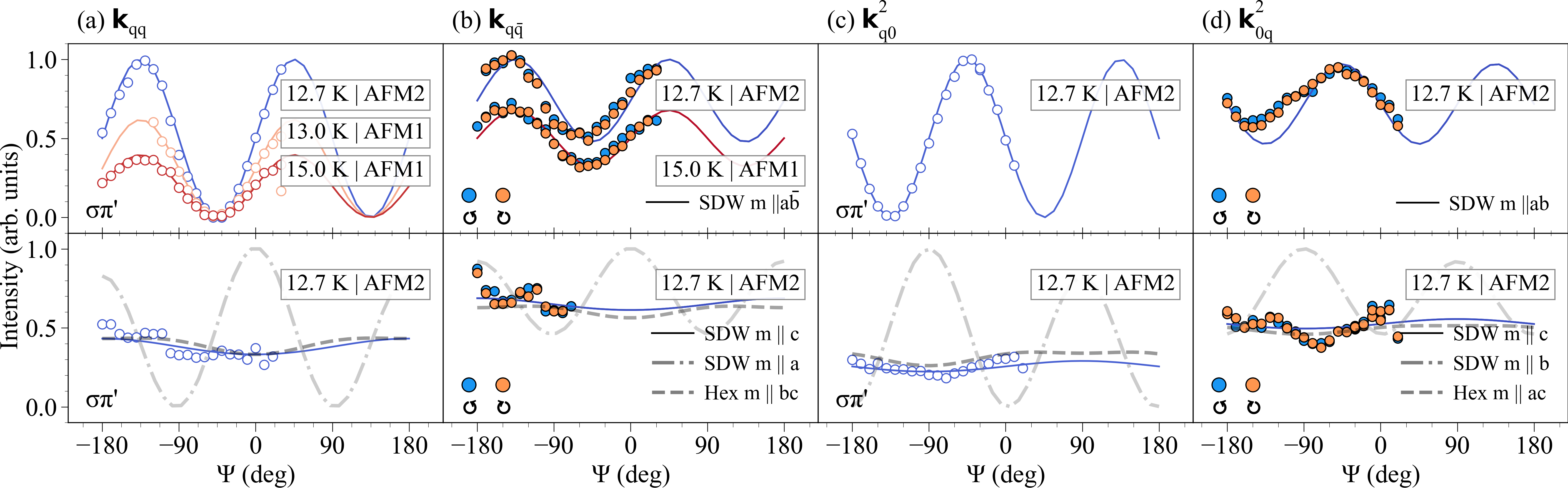}
\caption{\label{FIG::AFM2AZI}The scattered intensity from the \koneqq , \koneqmq , \ktwozq , \ktwoqz magnetic satellites of the (0,0,8) reflection as a function of azimuth collected in (a)-(d) $\sigma \pi '$ channel and (e)-(f) with incident CL ($\circlearrowleft$) or CR ($\circlearrowright$) polarised light in the AFM1 or AFM2 phases. The markers represent the normalised data, and the lines represent the fitted azimuthal dependencies for the different magnetic structure solutions.}
\end{figure*}

Resonant elastic x-ray scattering (REXS) measurements were performed at the I16 beamline at Diamond Light Source \cite{2010Collins} on an as grown 1 mm$^3$ sample of \ce{EuAl4}. The beamline was aligned at the Eu L3 edge, 6.972 keV. One of the key challenges that arises in determining the magnetic structure in the presence of multiple magnetic modulation vectors is to be able to distinguish whether these magnetic propagation vectors modulate the same magnetic phase, a multi-k structure, or whether each magnetic modulation vector corresponds to a distinct single-k domain that is separated spatially in the crystal \cite{2021Paddison}. In principle one can distinguish between these two scenarios if the incident beam on the crystal is smaller than the size of the domains, and thus by probing different positions on the sample, one may either observe a spatial separation of the various magnetic propagation vectors (single-k) or the existence of the magnetic propagation vectors at all positions on the sample (multi-k). In general the size of domains depends on a number of factors such as the thermal cycling of the sample, and local strain effects, so it is challenging to predict the expected size of the domains, and without a-priori knowledge of the domain size or distribution it becomes necessary to rely on other phenomenological and experimental methods to distinguish between the two scenarios \cite{2021Paddison}. According to Landau's theory of phase transitions \cite{1987Toledano}, for a multi-k structure one can expect the presence of free energy coupling terms that combine arms of the star of k with a secondary lattice modulation imposed by translation and time reversal symmetry. In this regard, one can search for these secondary distortions and the absence of such satellites is a good (but not definite) indication that the system is single-k. 

\section{Magnetic Structures}

In the first magnetically ordered phase, AFM1, present between T$_1$ and T$_2$, we observed two modulation vectors; \koneqq\ = (0.097(5),0.093(5),0) and \koneqmq\ = (0.092(5),-0.087(5),0), as shown in Fig. \ref{FIG::tempdep}(a), which lie along the DT line of symmetry ($\alpha$,$\alpha$,0). Below T$_2$, an additional two propagation vectors appeared, \ktwoqz\ = (0.175(5),0,0) and \ktwozq\ = (0,0.178(5),0). The associated reflections were of approximately equal intensity, but considerably weaker, by at least two orders of magnitude compared with those related to the \ktwozq\ and \ktwoqz\ propagation vectors. Below T$_3$, the intensity of reflections measured with propagation vectors \koneqq\ and \koneqmq\ rapidly reduced to zero. Concomitant with this was a large increase in scattering measured from reflections with propagation vector \kthreezq\ , where the magnitude of the propagation vector jumps from $0.175(5)$ to $0.165(5)$, as shown in Fig. \ref{FIG::tempdep}. Below T$_4$, \kfourzq\ changes from $0.165(5)$ to $0.188(5)$, whilst remaining along the same direction, with a noticeable increase in the intensity of the scattering. The above reflections were identified to be magnetic in origin from a measurement of the polarisation and energy dependence, shown in Fig. S2 of the SM. These observations are in agreement with the results of the single crystal neutron diffraction presented in Ref. \cite{2021Kaneko}, with the exception that we observed that the \ktwozq\ and \ktwoqz\ were also present in the AFM2 phase \footnote{It is likely that the single crystal neutron diffraction experiments were not able to detect these satellites owing to the large Eu neutron absorption cross section and the weak intensity of these peaks in the AFM2 phase.}. 

In order to probe the spatial arrangement of the modulation vectors in each of the phases, and thus attempt to clarify its multi-k vs single-k nature, the incident x-ray beam was focussed to 100$\micro$m by 37$\micro$m spot, and the response across the sample was measured in each of the magnetically ordered phases. To identify the orientation of the spins we performed azimuthal dependencies, as in Ref. \cite{2023Vibhakar}, using linearly polarised light. The azimuthal dependencies were repeated using incident circular polarised light to measure the collinearity of the spin structure. A contrast in the intensity of the magnetic satellites when probed with oppositely polarised circular light would occur if the spins are in a non-collinear arrangement, whilst an absence of any contrast implies the spin structure is collinear or equally populated by non-collinear inversion domains. 

\subsection{AFM1 phase}

In the AFM1 phase, the \koneqq\ and \koneqmq\ magnetic satellites of the (0,0,8) reflection were present across the entire sample, as shown in Fig. \ref{FIG::AFM12_structure}(a)-(b). Even though we did not observe any spatial segregation of the two satellites, we did observe that the ratio of their scattered intensity varied significantly across the sample, Fig. \ref{FIG::AFM12_structure}(c), which implies the magnetic structure was not multi-k, a multi-k structure would give rise to a constant ratio. Furthermore for a multi-k solution one would expect structural modulations with propagation vector \koneqq\ + \koneqmq\ \cite{2023Wood}, as this would couple the two magnetic propagation vectors in a trilinear free energy invariant that does not break translational symmetry, which was not observed. Together this suggests that the AFM1 phase is single-k, with domains that are smaller than the focussed beam spot. Azimuthal scans measured on the magnetic satellites of the \koneqq\ and \koneqmq\ are shown in Fig. \ref{FIG::AFM2AZI}(b)-(c). The absence of any contrast in the intensity of the satellites when probed with incident circular light, Fig. \ref{FIG::AFM2AZI}(c), suggests that the spin structure is collinear. The three possible spin structures determined from symmetry, details of which are given by Sec. S1 of the SM, were used to fit the data with a single variable parameter, and the only structure to give an excellent fit to the data, Fig. \ref{FIG::AFM2AZI}(b)-(c), was a spin density wave where the moments are aligned in the $ab$ plane perpendicular to the direction of the propagation vector, as shown in Fig. \ref{FIG::AFM12_structure}(h). 

\subsection{AFM2 phase}

\begin{figure*}[ht]
\centering
\includegraphics[width =\linewidth]{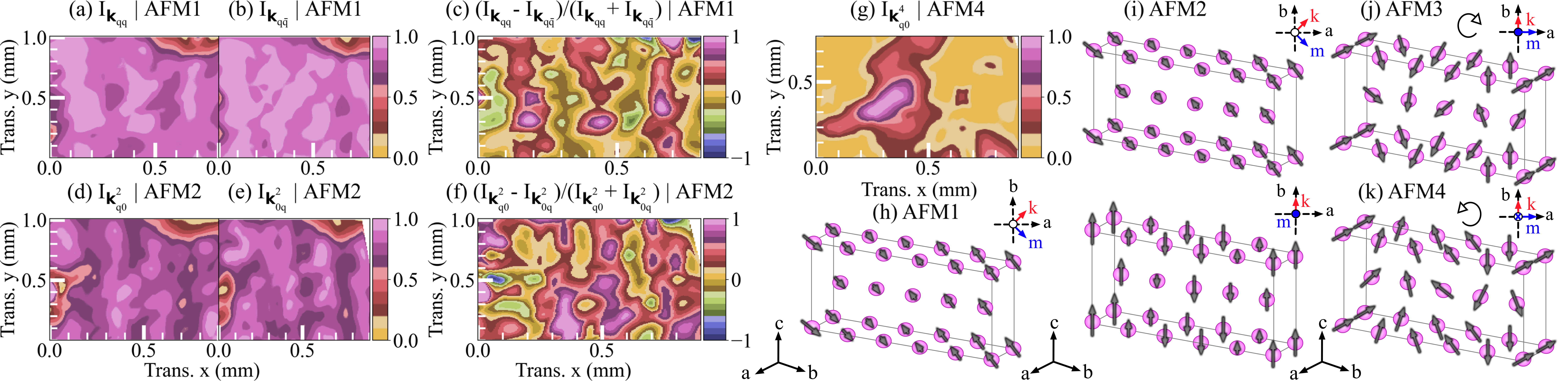}
\caption{\label{FIG::AFM12_structure}(a)-(g) The variation in intensity of the magnetic satellites of the (0,0,8) reflection taken as the beam was translated across the sample. (a) The \koneqq\ (b) \koneqmq\ and (c) the difference in intensity of the \koneqq\ and \koneqmq\ normalised by the sum of their intensity taken at 13.7 K in the AFM1 phase. (d) The \ktwoqz\ (e) \ktwozq\ and (f) the difference in intensity of the \ktwoqz\  and  \ktwozq  normalised by the sum of their intensity taken at 12.7 K in the AFM2 phase.(g) The $\mathbf{k}^4_{\mathrm{q0}}$ taken at 7 K in the AFM4 pfhase. A representation of the spin order stabilised in each of the four magnetically ordered phases is given for the $\mathbf{k}_{\mathrm{qq}}$ and $\mathbf{k}_{\mathrm{0q}}$ propagation vectors.}
\end{figure*}

In the AFM2 phase, the \koneqq , \koneqmq , \ktwoqz\ and \ktwozq\ were observed to exist at all points on the sample, Fig. \ref{FIG::AFM12_structure}(d)-(e). As  \ktwoqz\ and \ktwozq\ are equal to \koneqq\ $+$ \koneqmq\ and \koneqq\ $-$ \koneqmq\ respectively, within the experimental error, these could be the peaks that would appear in the presence of a multi-k structure. However as they are magnetic and not structural, it would necessitate the appearance of magnetism at the $\Gamma$-point so that the free energy term coupling these order parameters is fourth power, and therefore invariant upon time-reversal. Physically this would manifest as ferromagnetic order, which was not observed from magnetometry data \cite{2023Moya, 2015Nakamura} or from previous single crystal neutron diffraction studies \cite{2021Kaneko} indicating that together \koneqq\ and \koneqmq\ do not form a multi-k structure. Azimuthal scans measured on the magnetic satellites of the (0,0,8) show that the magnetic ordering associated with propagation vector \koneqq\ (and \koneqmq) was still a SDW m$||a\bar{b}$ (m$||ab$), as shown in Fig. \ref{FIG::AFM2AZI}(b)-(c). The azimuthal data collected on the \ktwoqz\ and \ktwozq\ satellites of the (0,0,8) reflection could be fit with two possible spin structures as shown in Fig. \ref{FIG::AFM2AZI}(d)-(e); a SDW m$||c$ with a single variable parameter or a helix with m$||bc$ (m$||ac$) with three variable parameters. In the case of the helix the data were fit assuming the presence of equal populations of both inversion domains, given the lack of contrast in the circular light, and by assuming a highly elliptical helix where the ratio of the fourier components F$_{y(x)}$/F$_z$ was between 0.2 and 0.3. If a helix is indeed stabilised in the AFM2 phase one would expect the symmetry to lower to monoclinic as discussed in Sec. \ref{SEC::Structure}, which we have no evidence for, favouring the SDW m$||c$ solution, which is illustrated in Fig. \ref{FIG::AFM12_structure}(i).

\subsection{AFM3 phase}

While the \ktwoqz\ and \ktwozq\ propagation vectors were approximately of equal intensity in the AFM2 phase, in the AFM3 phase the intensity of $\mathbf{k}^3_{\mathrm{q0}}$ was considerably reduced, by a factor of 5 compared with the \kthreezq\, and was not present across the entire sample, it could only be found in a 300 $\micro$m by 300 $\micro$m region, as shown in Fig. \ref{FIG::AFM12_structure}(g). Each of the magnetic structure solutions identified by symmetry, Sec. S1 of the SM, was used to fit the azimuthal data collected on the \kthreezq\ satellite of the (0,0,8) reflection, and an excellent fit was only achieved using a helical magnetic structure model with moments oriented in the $ac$ plane, as shown in Fig. \ref{FIG::AFM12_structure}(j). To fit the azimuthal data consisting of 66 data points, we used 2 parameters; the amplitude of the oscillation of the helix and its ellipticity. An excellent fit was achieved with an ellipticity of 1.4 (F$_x$/F$_z$), as shown in Fig. \ref{FIG::AFM3AFM4}(a).

\subsection{AFM4 phase}

\begin{figure}[ht]
\centering
\includegraphics[width =\linewidth]{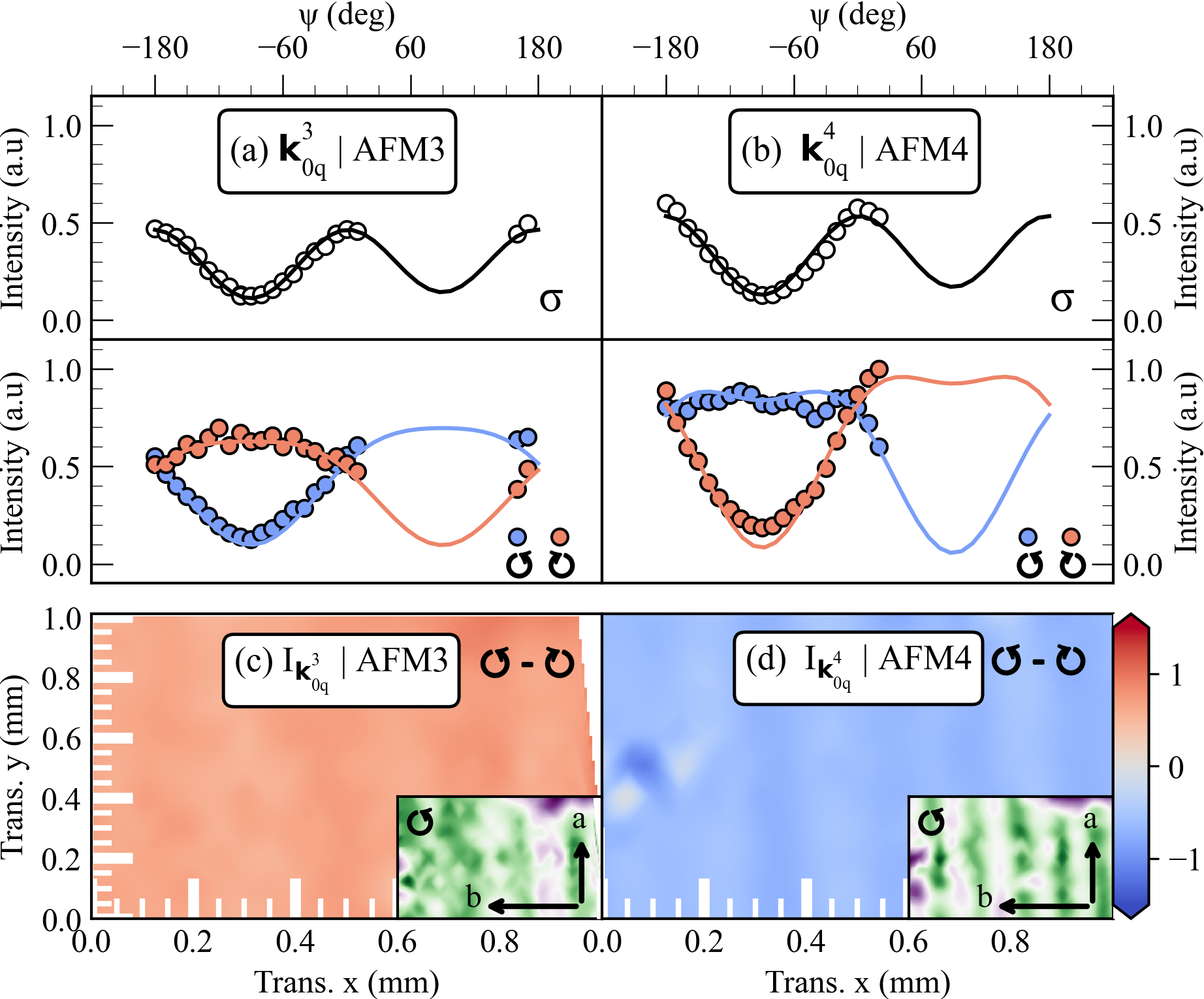}
\caption{\label{FIG::AFM3AFM4}The scattered intensity from the $\mathbf{k}_{\mathrm{0q}}$ satellite of the (0,0,8) reflection collected as a function of azimuth in (a) the AFM3 phase at 10.6 K and (b) the AFM4 phase at 7 K. The incident polarisation of light was either $\sigma$, CL ($\circlearrowleft$) or CR ($\circlearrowright$). The data, represented by the circular markers, are normalised against the data collected in the CR channel at 7 K. The solid lines represent the fit to the data using a helical magnetic structure with moments oriented in the $ac$ plane. Maps showing the variation in the difference in intensity of the $\mathbf{k}_{\mathrm{0q}}$ satellite of the (0,0,8) reflection at $\psi$ = -90 deg, collected with incident CL and CR light as the beam was translated across the sample at (c) 10.6 K in the AFM3 phase and (d) 7 K in the AFM4 phase. Insets show the same as the figure but plotting the response of the incident CL light and not the difference.}
\end{figure}

To fit the azimuthal scan of the \kfourzq\ satellite of the (0,0,8) reflection collected in the AFM4 phase, we repeated the procedure outlined above, and found that the best fit was still a helical magnetic structure solution with moments orientated in the $ac$ plane, but with a reversal of the chirality, such that now the best fit to the azimuthal data, Fig. \ref{FIG::AFM3AFM4}(b), was achieved with an ellipticity of -1.32 (F$_x$/F$_z$), as illustrated in Fig. \ref{FIG::AFM12_structure}(k).

As the x-ray beam does not probe the entire sample, we questioned whether we had measured a reversal in chirality or a shift in the magnetic chiral domain pattern that may occur with temperature. To clarify this matter we rastered the entire sample with the focussed beam spot, first using CR incident light, and then with CL polarised light at 7 K and 10.6 K, representative of the AFM4 and AFM3 phases respectively. We found there to be a small variation in the difference in intensity scattered from the circular light, but largely the sign of the difference was constant across the entire sample for a given temperature, as shown in Fig. \ref{FIG::AFM3AFM4} (c)-(d). Rather remarkably upon warming to the AFM3 phase, the sign of the difference flipped, demonstrating that the magnetic chirality across the whole sample had reversed upon warming!

\section{Magnetic Anisotropy}

For the rare earth ions, with the exception of Gd$^{3+}$ that has a half-filled 4$f$ sub-shell, anisotropic interactions typically arise from perturbative effects of the crystalline electric field on the free ion wavefunctions of the rare earth ions \cite{1952Stevens, 1953Elliot}, where L and S are evaluated using the Russell-Saunders coupling scheme. Eu$^{2+}$, which is isoelectronic to Gd$^{3+}$, also has S = 7/2 and L = 0, and is thus expected to be isotropic with respect to its crystallographic environment. However the adoption of anisotropic magnetic structures clearly points to the presence of anisotropy. One could make the argument that the assumption of a fully quenched orbital angular momentum may not be true, and hence there may be a small finite 4$f$ spin-orbit coupling interaction, however it would not explain the temperature dependent change in anisotropy. Instead we adopt the theory used to explain the magnetocrystalline anisotropy present in elemental Gd \cite{2003Tosti}.

Treating the 4$f$ states as a spin-polarised core that is well localised, and the 5$d$ and 6$s$ as the hybridised valence conduction bands, which from angle-resolved photoelectron spectroscopy measurements \cite{2016Kobata} and numerous first principles calculations \cite{1972Elliott, 2003Tosti} has show to be a good assumption, it was shown that the calculated easy axis in elemental Gd is indirectly dependent on the magnitude of the magnetic moment developed on the 4$f$ sites \cite{2003Tosti, 2005Tosti}. The exchange field from the 4$f$ ions can split the 5$d$-6$s$ conduction bands, giving rise to a valence band moment, and a spin-orbit interaction from the conduction electrons that can act as a source of magnetocrystalline anisotropy \cite{2003Tosti}. The spin-orbit coupling is directly dependent on the exchange splitting of the conduction band, which in turn is dependent on the magnitude of the 4$f$ moment. Hence, a change in the ordered 4$f$ moment can dramatically change the spin-orbit coupling, and therefore the magnetic anisotropy. While no calculations have been performed for Eu$^{2+}$, this source of magnetic anisotropy is expected to be present for all rare earths, but is never observed owing to the dominance of crystal electric field effects for cases where L$\neq 0$ \cite{2003Tosti}. In \ce{EuAl4} the valence bands are expected to be dominated by the Eu atomic orbitals \cite{2016Kobata}, and hence is not expected to change the results above. 

\subsection{Dipolar Calculations}

\begin{figure}[ht]
\centering
\includegraphics[width =\linewidth]{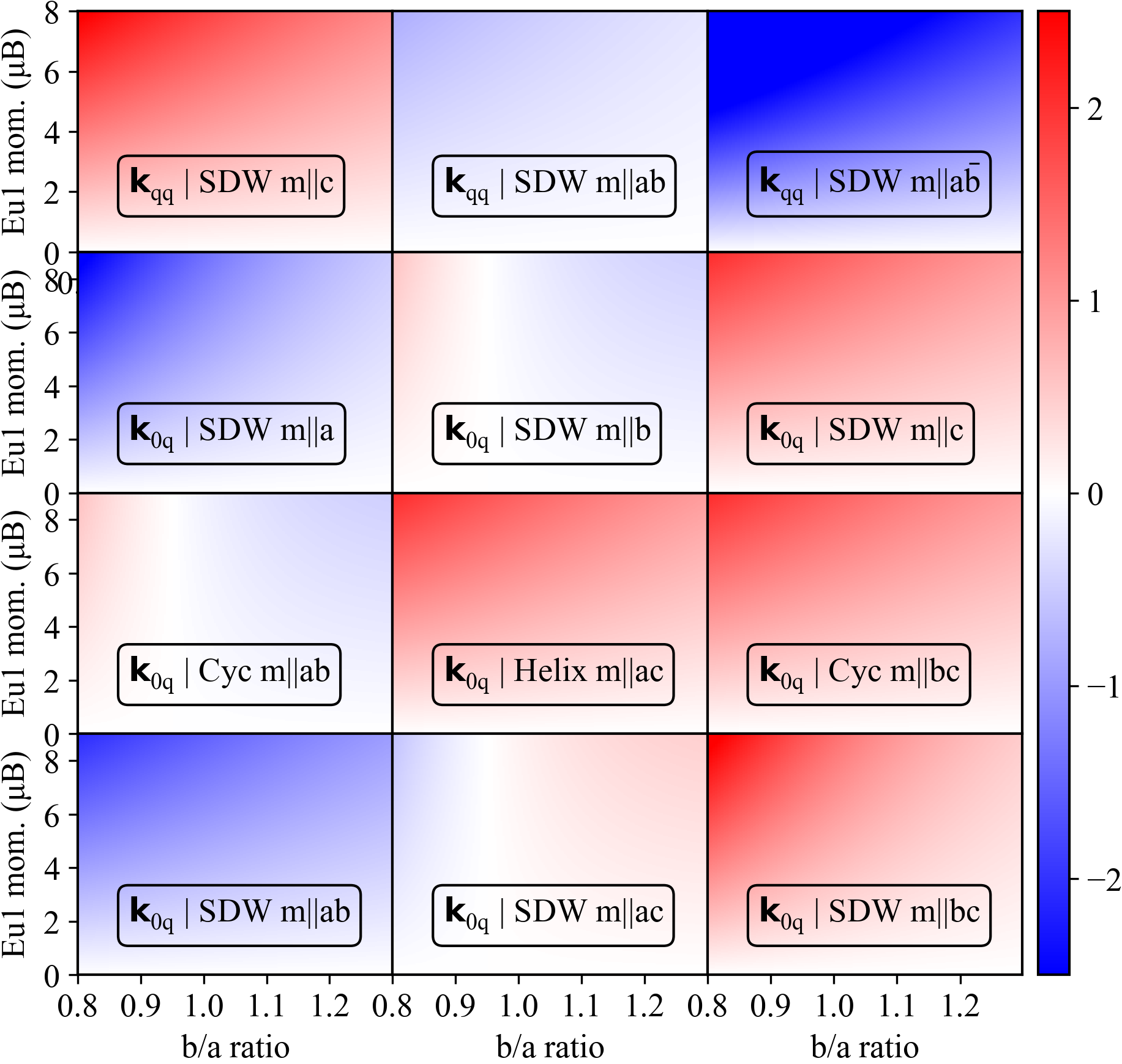}
\caption{\label{FIG::dipole} The dipole energy of \ce{EuAl4} calculated for the Eu1 ion taking into account all Eu ions in two unit cells along each crystallographic direction, as a function of the magnitude of the Eu1 moment and the ratio of the $b$:$a$ lattice parameters, for each of the symmetry allowed spin configurations for the \koneqq\ and $\mathbf{k}_{\mathrm{0q}}$ propagation vector. The calculations were evaluated for \koneqq\ = (0.09,0.09,0) and $\mathbf{k}_{\mathrm{0q}}$ = (0,0.19,0).}
\end{figure}

Besides the spin orbit coupling from the 5$d$-6$s$ bands, the 4$f$ dipolar interactions are also expected to contribute to the magnetic anisotropy \cite{2003Tosti}. We emphasise that the dipolar interaction is unlikely to drive the ordering of the Eu ions in \ce{EuAl4} as the dipolar energy is on the order of $\sim$ 1 K, however it can be relevant in choosing the magnetic anisotropy in the material, given that no dominant sources of anisotropy are expected to be present. We calculated the dipolar energy, given in Eqn. \ref{EQN::dipolar} \cite{1999Jackson}, for each of the different spin structure solutions, whilst varying both the magnitude of the Eu moment and the ratio of the $a/b$ lattice parameters, and the results are shown in Fig. \ref{FIG::dipole}. In performing these calculations we assumed that the Eu ions do not contribute to the periodic distortion of the lattice that gives rise to the CDW satellites, in accordance with Ref. \cite{2021Kaneko, 2022Ramakrishan} that suggests it is the Al ions that displace. We note changing the magnitude of \textbf{k}$_{\mathrm{0q}}$ did not change the qualitative results of the calculations, and hence these results are valid for the AFM2, AFM3 and AFM4 phases.  

\begin{equation}
\mathrm{E}_\mathrm{dip} = \frac{\mu_0}{4\pi r^3}[\mu_1 \cdot \mu_2 - \frac{3}{r^2}(\mu_1 \cdot r)(\mu_2 \cdot r)]
\label{EQN::dipolar}
\end{equation}

Fig. \ref{FIG::dipole} shows that for the \koneqq\ propagation vector, the spin configuration that minimises the dipolar energy is a SDW with the moments orientated perpendicular to the direction of the propagation vector in the $ab$ plane, i.e for m$||a\bar{b}$, as observed experimentally. For the $\mathbf{k}_{\mathrm{0q}}$ propagation vector, the spin configuration that minimises the dipolar energy is a SDW $m||a$. These calculations show that the dipole-dipole interaction selects the moment in the $ab$ plane to be orientated perpendicular to the direction of the propagation vector, in agreement with our experimental observations. Whilst dipolar interactions are sufficient to explain the observed anisotropy of the spin structures associated with the \koneqq\ propagation vector in the AFM1 and AFM2 phases, it cannot explain why an additional component along $c$ is favoured for the $\mathbf{k}_{\mathrm{0q}}$ propagation vector, indicating other anisotropic terms become energetically favourable as the temperature is lowered below T$_3$, such as those discussed in the preceding paragraph. The dipolar interaction is dependent on the magnitude of the magnetic moments and the distance between them, and is therefore to applicable to other systems that order in the same symmetry and with similar sized moments, such as the square-net systems \ce{GdRu2Si2} \cite{2020Khanh} and \ce{EuGa2Al2} \cite{2023Vibhakar}, where the in-plane moment has indeed been observed to order perpendicular to the direction of the propagation vector, further validating our calculations.

\begin{figure*}[ht]
\centering
\includegraphics[width =\linewidth]{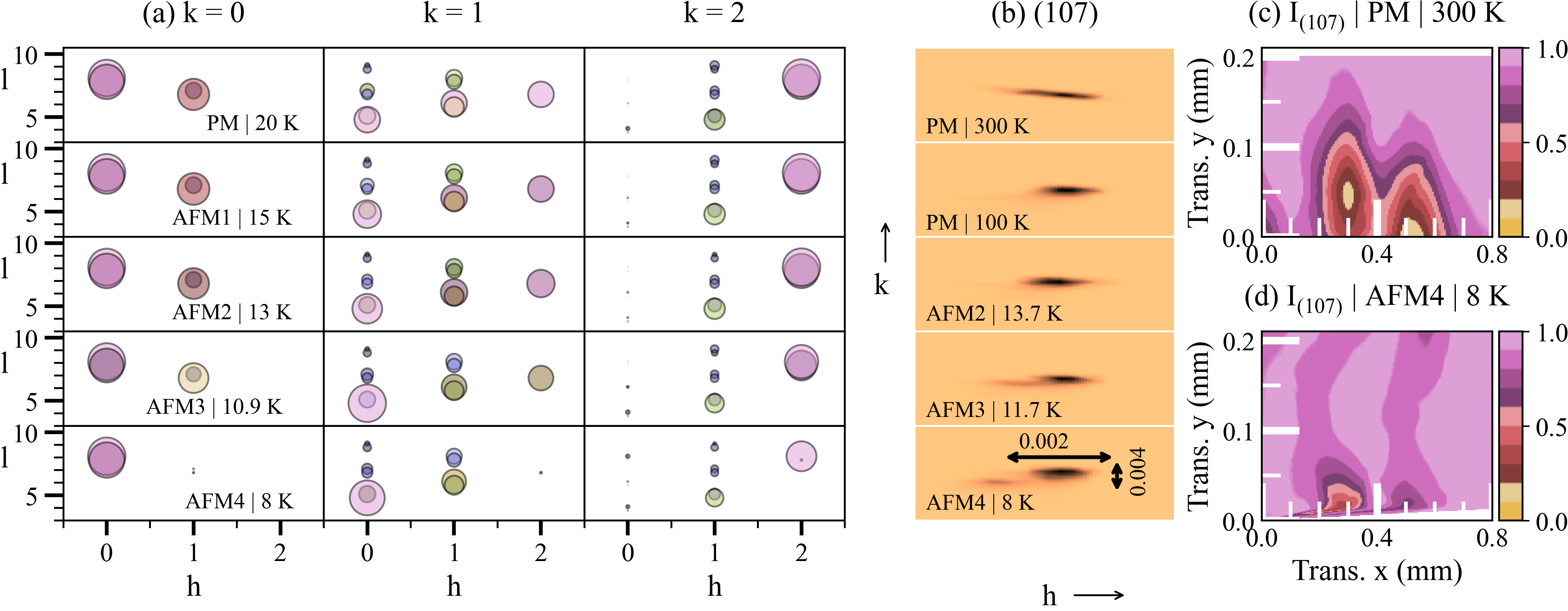}
\caption{\label{FIG::chargeCDWdomains} (a) Variation in the intensity of CDW reflections (h,k,l+$\tau$) as the sample was warmed through each magnetically ordered phase. The intensity of the reflection is represented by both the size and colour of the marker. (b) Reciprocal space maps showing the (1,0,7) reflection in the $hk$ plane as the sample was warmed. Maps of the intensity of the (1,0,7) reflection collected as the beam was translated across the sample at (c) 300 K in the PM phase and (d) at 8 K in the AFM4 phase.}
\end{figure*}

\section{Structural and symmetry changes}\label{SEC::Structure}

Below T$_3$ we observed a change to the average crystal structure, from a splitting of the (1,0,7) reflection in the $hk$ plane as shown in Fig. \ref{FIG::chargeCDWdomains}. No further discontinuous changes to the Bragg reflections were observed below T$_4$, however large changes to the intensity of the CDW satellites were observed. To measure the CDW reflections we realigned the beamline to 6.5 keV, just below the Eu L3 edge, so that we could only be sensitive to changes associated with the ionic displacements of the CDW order. The intensity of the CDW reflections underwent small shifts between 20 K (PM) and T$_3$, and larger changes below T$_4$. Of particular importance were reflections such as the (1,0,7+$\tau$) and (2,1,7+$\tau$), that had finite intensity in all magnetically ordered phases, but went to zero in the AFM4 phase, Fig. \ref{FIG::chargeCDWdomains}(a). A refinement of the CDW structure was not possible from the data collected, and hence neither was a complete determination of the space group symmetry. 

Nevertheless from a phenomenological point of view, an asymmetry between two spin chiral states would not occur unless the crystal structure itself was chiral or polar. Indeed an analysis of the symmetry of the magnetic and nuclear structures suggests that the symmetry is lowered to a polar monoclinic space group below T$_3$. The structure of the CDW below T$_\mathrm{CDW}$ is reported to be described by the $Fmmm(00\gamma)s00$ \cite{2022Ramakrishan} or by the $Immm(00\gamma)s00$ \cite{2024Korshunov} superspace groups. Both these subgroups derive from the $I4/mmm$ parent structure from the $\Lambda$5 irreducible representation (irrep) with different order parameter directions (ODP) ($a,a,a,-a$) and ($a,a,0,0$) respectively. The magnetic helical structure, which is present in the AFM3 and AFM4 phases, is described by the action of two irreps of the parent $I4/mmm$ structure, namely $m\Sigma_3$ and $m\Sigma_4$. Considering only the magnetic degree of freedom, the obtained magnetic superspace group is $I222.1’(00\gamma)00ss$, which reflects the chiral nature of the helical phase. To obtain the total symmetry of the AFM3 and AFM4 phases is necessary to combine the symmetry requirement of both the CDW and helical phases. This will result in a (3+2)D magnetic superspace group able to correctly describe the system symmetry. Since there is an uncertainty regarding the order parameter direction of the CDW phase is necessary to consider both reported structures. If we consider the $Fmmm(00\gamma)s00$ structure reported in \cite{2022Ramakrishan}, the resulting magnetic superspace group is $B2.1’(\alpha_1,\beta_1,0)0s(0,0,\gamma_2)s0$. This magnetic superspace groups will lead to a change of the OPD of the CDW distortion that is consistent with the observation of changes in the CDW peaks. The monoclinic strain allowed in this symmetry group will give rise to two strain domains in the same chiral magnetic domain, as has been observed in numerous square net systems \cite{2000Gourdon, 2006Hucker, 2016Zhang, 2021Wu, 2022Frison}, which can explain the observed stripe contrast observed in the inset of Fig. \ref{FIG::AFM3AFM4} and Fig. \ref{FIG::chargeCDWdomains}(d), as well as explaining the nuclear reflection splitting shown in Fig. \ref{FIG::chargeCDWdomains}(b). The situation is slightly different if the $Immm(00\gamma)s00$ superspace group is considered as the symmetry of the CDW above T$_3$. In this case there are two possible candidates for the final symmetry depending on if the order parameter direction of the CDW changes or not. In the case where the CDW symmetry remains the same, the $I222.1’(0,\beta_1,0)0s0s(0,0,\gamma_2)00s0$ superspace group will describe the final state. This superspace group cannot explain the observed splitting of the nuclear reflection below T$_3$, nor the stripe contrast. On the contrary, if the ODP of the CDW changes then $B2.1’(\alpha_1,\beta_1,0)0s(0,0,\gamma_2)s0$ is the correct superspace group. The latter case is the most likely since it can explain the stripe contrast observed in Fig. \ref{FIG::AFM3AFM4} and Fig. \ref{FIG::chargeCDWdomains}, as also for this case in a single magnetic chiral domain we should expect two monoclinic strain domains. It is worth noticing that in both monoclinic cases the symmetry of the CDW (without considering the magnetic degrees of freedom) is now chiral having point group 222 or 2. 

We further note that in inversion symmetric crystals one expects to establish equal equally populated chiral magnetic domains that restores the global parity of the crystal \cite{2008Marty}, whilst for non-centrosymmetric crystals an asymmetry in the chiral domain population is expected, and often stabilises a single magnetic chiral domain \cite{1983Shirane, 2008Marty, 2010Janoschek}. Hence, the observation of a single chiral magnetic domain in \ce{EuAl4} is further evidence of the breaking of inversion symmetry in this crystal below T$_3$. This change in the CDW symmetry at T$_3$, and its now chiral nature is important in explaining the chirality reversal observed at T$_4$ as it will be discussed in the next section.

\section{Reversal of the spin chirality}

The establishment of a polar symmetry below T$_3$ can break the degeneracy between the two chiral states of the magnetic helix. Without the chiral or polar structural distortions, the two chiral states of the magnetic helix represent degenerate chiral domains. A change to the polarity or chirality of the crystal structure could then induce a change to the magnetic chirality, if the magnetic and structural order parameters were coupled \cite{2023Vibhakar}. It is also possible that we observe a new type of phase transition with a spontaneous change of magnetic chirality, but this scenario again requires the asymmetry between the two chiral state of the magnetic helix and therefore the chiral or polar crystal structure. An alternative mechanism to reverse the spin chirality is the competition between Dzylaoshinskii Moriya and dipolar interactions tuned by RKKY interactions, which has been shown to reverse the chirality of domain walls \cite{2019Lucassen}.

\section{Conclusions}

To summarise, we demonstrate a spontaneous reversal of spin chirality in a single crystal sample of the intermetallic magnet \ce{EuAl4}. In the AFM1 phase \ce{EuAl4} orders to form a SDW with moments aligned in the $ab$ plane, perpendicular to the propagation vector. The AFM2 phase is characterised by the coexistence of two spin orders, the SDW present in the AFM1 phase, and a SDW with m$||c$. In the AFM3 phase the ordering of the magnetic modulations change, and a magnetic helix of a single chiral domain is stabilised, where one component of the moment is in the $ab$ plane, perpendicular to the propagation vector, and the second is along $c$. We find that in every magnetically ordered phase of \ce{EuAl4} the in-plane moment is perpendicular to the orientation of the magnetic propagation vector, which we demonstrate is favoured by magnetic dipolar interactions. Below T$_4$ the magnetic chirality of the helix reverses, and the sample remains as a single chiral domain. Concomitant with the establishment of the helical magnetic ordering is the lowering of the crystal symmetry to monoclinic, as evidenced the formation of uniaxial charge and spin strip domains. A group theoretical analysis of the nuclear and magnetic structures demonstrates that below T$_3$, the symmetry lowers to polar monoclinic, explaining the observed asymmetry in the chiral states of the magnetic helix. We find that the change in the CDW symmetry at T$_3$ and its now chiral nature is necessary to explain the observed spin chiral reversal.
\section{Acknowledgments}
The authors gratefully acknowledge the technical support provided at the I16 Beamline at Diamond Light Source to perform the resonant x-ray scattering measurements (Exp. MM31813-1 and Exp. NR35465-1).

\section{Appendix}

\section{Resonant Elastic X-ray Scattering} \label{SEC::RXSexp}
Resonant x-ray scattering (RXS) measurements were performed in the vertical scattering geometry with an incident beam size of 180 $\micro$m in the horizontal direction and $\sim$30 $\micro$m in the vertical. The sample was mounted so that the (0,0,$l$) was specular, and the azimuthal reference was ($h$,0,0), such that at $\phi$ = 0 the $a$ axis was parallel to the incident beam. An ARS closed cycle refrigerator was used to cool the sample from room temperature to 5 K. A Pilatus 100K area detector was used to measure the diffracted intensity when it was not analysed into the $\sigma$ and $\pi$ channels. To analyse the diffracted beam a Cu (220) crystal orientated close to 44.66$\degree$ with respect to the incident beam was used together with a Merlin area detector. We note that at 6.972 keV the attenuation length of the x-rays is expected to be approximately 10$\micro$m for a grazing incident angle greater than 40$\degree$, which serves as an estimate for the x-ray penetration depth.

\subsection{Polariser}

The incident polarisation of light was tuned between linear horizontal ($\sigma$), linear vertical ($\pi$) and two circular polarisations of opposite handedness, circular right ($\circlearrowright$) and  circular left ($\circlearrowleft$), using the light transmitted through a 1 mm diamond (0,0,$1$) plate orientated close to the (1,1,1) reflection in the asymmetric Laue mode.

We note that the conversion of the light from linear vertical to circular is not perfect, and thus a fraction of the light remains $\sigma$ polarised. Polarised light can be described by the three real Stokes parameters, P$_1$, P$_2$ and P$_3$, which form the vector \textbf{P} as in Ref. \cite{1993Lovesey}. For perfectly circular polarised light $ \mathbf{\mathrm{P}}=(0,\pm 1,0)$. We fitted the following expression, $ \mathbf{\mathrm{P}}=(0,\mathrm{P}_2,\sqrt{1-\mathrm{P}_2^2})$, which assumes the beam is partially circular polarised and partially linearly polarised, and determined the value of P$_2$ to be 0.88. 

\subsection{Resonant magnetic x-ray scattering}

$\eta$ scans measured in the $\sigma\sigma '$ and $\sigma \pi '$ channels and an energy dependence of the magnetic satellites of the (0,0,8) reflection in the AFM1, AFM2 and AFM4 phases is shown below. An absence of scattering in the $\sigma\sigma '$ channel and the presence of scattering in the $\sigma \pi '$ confirms the magnetic origin of the intensity at these satellite reflections.

\begin{figure}[ht]
\centering
\includegraphics[width =\linewidth]{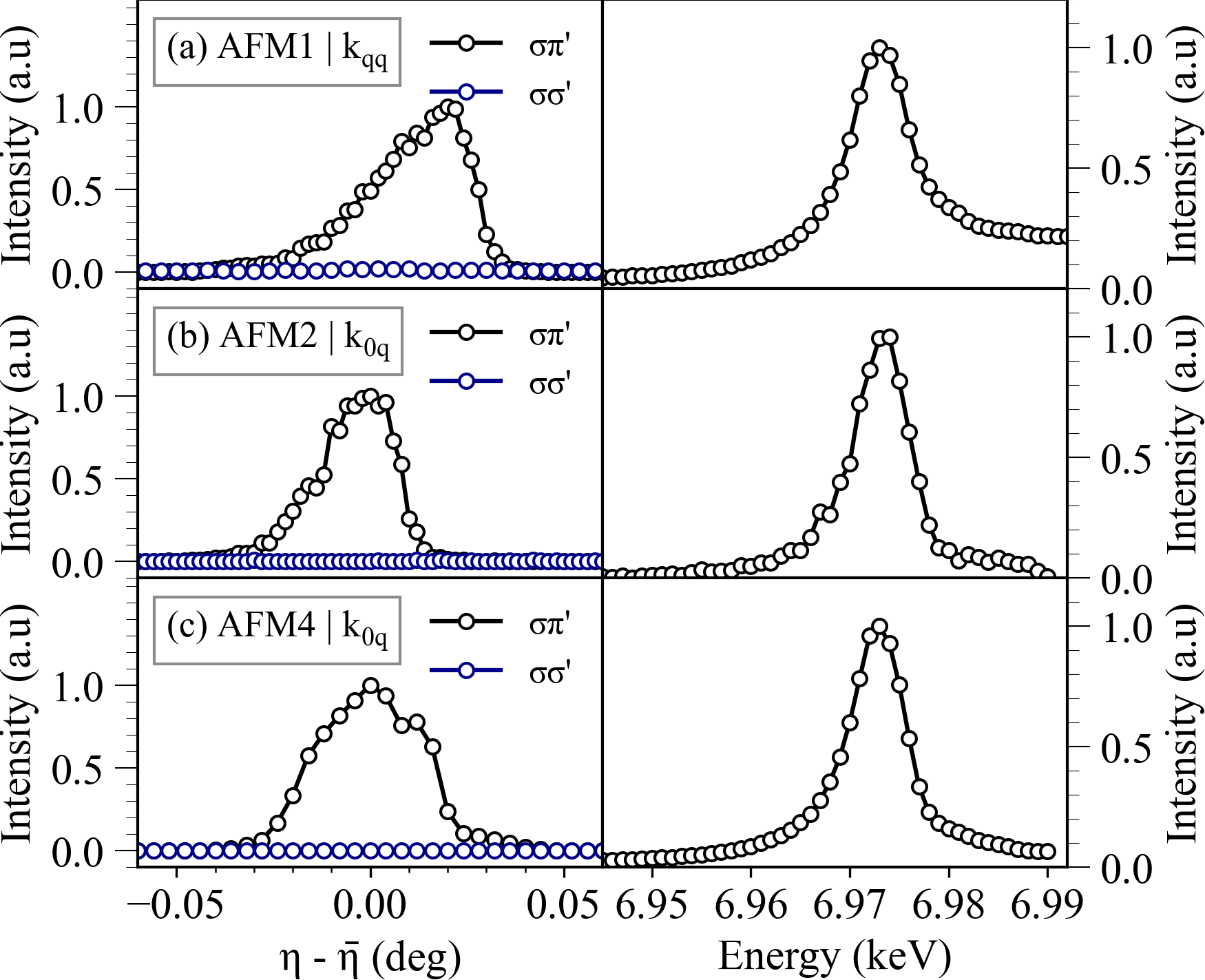}
\caption{\label{FIG::SMenergypol}$\eta$ scans collected on the magnetic satellites of the (0,0,8) reflection in the $\sigma\sigma$' and $\sigma\pi$' channels and energy scans of the (0,0,8) reflection collected in (a)-(b) the AFM1 phase at 13 K (c)-(d) in the AFM2 phase at 12.7 K and (e)-(f) in the AFM4 phase at 8 K.}
\end{figure}
\section{Symmetry analysis}\label{SEC::symmetry}

To determine the magnetic structures predicted by symmetry, the space group of the parent crystal structure was maintained as $I4/mmm$. Whilst the space group symmetry is likely to be lower, given the onset of the CDW precedes the condensation of the magnetic order, one can use a higher symmetry parent space group when performing the analysis, especially since a determination of the crystal structure was not performed below T$_{\mathrm{CDW}}$. The symmetry analysis was performed using \textsc{ISODISTORT} from the \textsc{ISOTROPY} Software Suite \cite{HStokes, 2006Campbell}. 

The magnetic representation of the Wyckoff positions of the Eu11 and Eu12 ions in the $I4/mmm$ symmetry for the incommensurate propagation vector, ($\alpha$,$\alpha$,0), decomposes into three irreducible representations, each of which corresponds to a distinct ordering of the Eu spins, as summarised in Table. \ref{TAB::aa0singleirrepsol}. The three possible spin orderings are spin density waves, where the direction of the moments are either $||c$, $||ab$ or $||a\bar{b}$. An analysis of the irreducible representations of the magnetic representation for the Wyckoff positions of the Eu11 and Eu12 ions in the $I4/mmm$ symmetry for the  the incommensurate propagation vector, ($\alpha$,0,0), can be found in Ref. \cite{2023Vibhakar}, and hence we do reproduce the analysis here. Instead we summarise the possible spin orderings in Table \ref{TAB::a00singleirrepsol} and Table \ref{TAB::a00doubleirrepsol}.

\begin{table*}[h]
\setlength{\tabcolsep}{4.pt}
\begin{ruledtabular}
{\renewcommand{\arraystretch}{1.3}
\begin{tabular}{c c c}
Mode & Eu11 & Eu12\\
\hline
$\zeta_i^1$& $\exp(-2 \pi i (\mathbf{k} \cdot (0,0,0)) )$ & $\exp(-2 \pi i (\mathbf{k} \cdot (\frac{1}{2},\frac{1}{2},\frac{1}{2})) )$ \\
$\zeta_i^2$& $\exp(-2 \pi i (\mathbf{k} \cdot (0,0,0)) - \pi/2)$ & $\exp(-2 \pi i (\mathbf{k} \cdot (\frac{1}{2},\frac{1}{2},\frac{1}{2}))-\pi/2)$\\
\end{tabular}
}
\end{ruledtabular}
\caption{\label{TAB::magneticmodes}Symmetry adapted basis modes $\zeta_i^1$ and $\zeta_i^2$ describing the magnitude of the Fourier components of the magnetic moment for the Eu11 and Eu12 sites with fractional coordinates of (0,0,0) and ($\frac{1}{2}$,$\frac{1}{2}$,$\frac{1}{2}$) respectively with respect to the $I4/mmm$ unit cell. The subscript $i$ denotes the direction of the magnetic Fourier component, which can be along x ($||a$), y ($||b$) or z ($||c$), $\mathbf{k}$ represents the propagation vector of the magnetic ordering, and $\mathbf{r_i}$ the fractional coordinates of the atomic site with respect to the $I4/mmm$ unit cell.}
\end{table*}

\begin{table*}[h]
\setlength{\tabcolsep}{3.8pt}
\begin{ruledtabular}
{\renewcommand{\arraystretch}{1.3}
\begin{tabular}{c c c c c}
Irrep  & OPD & Magnetic modes &  $\mathbf{S}_k$ & Magnetic structure\\
\hline
\multirow{1}{*}{$m\Delta_2$} & ($a,b;0,0$) & $\zeta_z^1$, $\zeta_z^2$ & (0,0,$m$) & spin density wave \\
\hline
\multirow{1}{*}{$m\Delta_3$}  & ($a,b;0,0$) & $\zeta_{x \bar{y}}^1$, $\zeta_{x \bar{y}}^2$& ($m$,$-m$,0) &  spin density wave\\
\hline
\multirow{1}{*}{$m\Delta_4$} & ($a,b;0,0$) &$\zeta_{xy}^1$, $\zeta_{xy}^2$ & ($m$, $m$,0) &  spin density wave\\
\end{tabular}
}
\end{ruledtabular}
\caption{\label{TAB::aa0singleirrepsol}The magnetic representations of the ($\alpha$,$\alpha$,0) propagation vector described for the Eu ions of the $I4/mmm$ unit cell in terms of the symmetry adapted basis modes defined in Tab. \ref{TAB::magneticmodes}. Details of the order parameter direction (OPD), the direction of the magnetic Fourier components ($\mathbf{S}_k$), and the magnetic structure described by each of the irreps is also given.} 
\end{table*}

\begin{table*}[h]
\setlength{\tabcolsep}{3.8pt}
\begin{ruledtabular}
{\renewcommand{\arraystretch}{1.3}
\begin{tabular}{c c c c c}
Irrep  & OPD &  Magnetic modes  & $\mathbf{S}_k$ & Magnetic structure\\
\hline
\multirow{2}{*}{$m\Sigma_2$} & ($a,0;0,0$) & $\zeta_x^1$ & ($m$,0,0) & spin density wave \\
& ($a,b;0,0$) & $\zeta_x^1$, $\zeta_x^2$ & ($m$,0,0) &  spin density wave \\
\hline
\multirow{2}{*}{$m\Sigma_3$}  & ($a,0;0,0$) & $\zeta_y^1$ & (0,$m$,0) &  spin density wave\\
&  ($a,b;0,0$) & $\zeta_y^1$, $\zeta_y^2$  & (0,$m$,0) & spin density wave \\
\hline
\multirow{2}{*}{$m\Sigma_4$} & ($a,0;0,0$) & $\zeta_z^1$ & (0,0,$m$) &  spin density wave\\
 & ($a,b;0,0$)& $\zeta_z^1$, $\zeta_z^2$ & (0,0,$m$) &  spin density wave\\
\end{tabular}
}
\end{ruledtabular}
\caption{\label{TAB::a00singleirrepsol}The magnetic representations of the ($\alpha$,0,0) propagation vector described for the Eu ions of the $I4/mmm$ unit cell in terms of the symmetry adapted basis modes defined in Tab. \ref{TAB::magneticmodes}. Details of the order parameter direction (OPD), the direction of the magnetic Fourier components ($\mathbf{S}_k$), and the magnetic structure described by each of the irreps is also given.} 
\end{table*}

\begin{table*}[h]
\setlength{\tabcolsep}{3.8pt}
\begin{ruledtabular}
{\renewcommand{\arraystretch}{1.3}
\begin{tabular}{c c c c c}
Irreps & OPD & Magnetic modes & $\mathbf{S}_k$ & Magnetic structure\\
\hline
\multirow{2}{*}{$m\Sigma_2$ \& $m\Sigma_3$} & ($a,0;0,0$) & $\zeta_x^2$, $\zeta_y^1$ & ($m_x$,$m_y$,0)& spin density wave\\
& ($a,0;0,0 | b,0;0,0$) & $\zeta_x^1$, $\zeta_y^1$ & ($m_x$,$m_x$,0)& cycloid\\
\hline
\multirow{2}{*}{$m\Sigma_2$ \& $m\Sigma_4$} & ($a,0;0,0 | 0,b;0,0$) & $\zeta_x^2$, $\zeta_z^1$ & ($m_x$,0,$m_z$)& spin density wave\\
& ($a,0;0,0 | b,0;0,0$) & $\zeta_x^1$, $\zeta_z^1$ & ($m_x$,0,$m_x$)& cycloid\\
\hline
\multirow{3}{*}{$m\Sigma_3$ \& $m\Sigma_4$} & ($a,0;0,0 | 0,b;0,0$) & $\zeta_y^1$, $\zeta_z^2$ &  (0,$m_y$,$m_z$)& spin density wave\\
& ($a,0;0,0|b,0;0,0$) & $\zeta_y^1$, $\zeta_z^1$ & (0,$m_y$,$m_y$)& helix\\
\end{tabular}
}
\end{ruledtabular}
\caption{\label{TAB::a00doubleirrepsol}Symmetry allowed magnetic structure solutions constructed by combining the symmetry allowed basis modes from two different irreducible representations of the ($\alpha$,0,0) magnetic propagation vector of the Eu ions in the $I4/mmm$ unit cell (see Table \ref{TAB::a00singleirrepsol}). Details of the order parameter direction (OPD), the direction of the magnetic Fourier components ($\mathbf{S}_k$), and the magnetic structure described by combinations of irreps is also given.}
\end{table*}

\bibliographystyle{apsrev4-2}

\end{document}